\documentclass[letterpaper, 11pt]{article}
\usepackage[T1]{fontenc}
\usepackage{lmodern}
\usepackage{listings}

\usepackage{graphicx}
\usepackage{authblk}
\usepackage{amsmath}
\usepackage{geometry}
\usepackage{parskip}
\usepackage{setspace}
\usepackage{caption}
\usepackage{xcolor}
\usepackage{hyperref}
\usepackage{booktabs}
\usepackage{comment}
\usepackage{afterpage}
\usepackage{chemformula}
\usepackage{adjustbox}
\usepackage{pifont}
\usepackage{multirow}

\newcommand{\fplus}{\mathord{\text{\ding{58}}}}
\usepackage[style=ieee]{biblatex}

\AtEveryBibitem{\clearfield{issn}} 
\AtEveryBibitem{\clearfield{isbn}} 
\DeclareFieldFormat{url}{\url{#1}} 
\DeclareFieldFormat{doi}{\url{https://doi.org/#1}} 

\hypersetup{
    colorlinks=true,
    linkcolor=black,   
    urlcolor=teal,
    citecolor=blue
}

\title{\textbf{Density Functional Theory ToolKit (DFTTK) to Automate First-Principles Thermodynamics via the Quasiharmonic Approximation}}
\author{\small Nigel Lee En Hew\textsuperscript{1,*}, Luke Allen Myers\textsuperscript{1}, Axel van de Walle\textsuperscript{2}, Shun-Li Shang\textsuperscript{1}, Zi-Kui Liu\textsuperscript{1}}
\affil{\small \textsuperscript{1}Department of Materials Science and Engineering, The Pennsylvania State University, University Park, Pennsylvania 16802, USA}
\affil{\small \textsuperscript{2}School of Engineering, Brown University, Providence, RI 02912, USA}
\affil{\small \textsuperscript{*}Corresponding authors: njh5724@psu.edu}

\date{} 

\begin{document}

\maketitle

\vspace{-2em}  

\section*{Abstract}
The Helmholtz energy is a key thermodynamic quantity representing available energy to do work at a constant temperature and volume. Despite a well-established methodology from first-principles calculations, a comprehensive tool and database are still lacking. To address this gap, we developed an open-source Density Functional Theory Tool Kit (DFTTK), which automates first-principles thermodynamics using the quasiharmonic approximation (QHA) for Helmholtz energy predictions. This Python-based package provides a solution for automating the calculation and analysis of various contributions to Helmholtz energy, including the static total energy contributions at 0 K in terms of DFT-based energy-volume curves, vibrational contributions from the Debye-Grüneisen model and phonons, and thermal electronic contributions via the electronic density of states. The QHA is also implemented to calculate the Gibbs energy and associated properties at constant temperature and pressure. The present work demonstrates DFTTK's capabilities through case studies on a simple FCC \ch{Al} and various collinear magnetic configurations of Invar \ch{Fe3Pt}, where DFTTK enumerates all unique configurations and their associated multiplicities. DFTTK is freely available on GitHub, and its modular design allows for the easy addition of new workflows.

\newpage
\section{Introduction}
Density functional theory (DFT) is widely used to calculate the Helmholtz energy for a given configuration $k$, $F^k (V, T)$ \cite{Wang2004ThermodynamicCalculations, Shang2010First-principlesNi3Al, Shang2024Revisiting625}:

\begin{equation}
F^k (V,T) = E^{k,0} (V) + F^{k,{vib}} (V,T)+ F^{k,{el}} (V,T),
\label{eq:1}
\end{equation}

where $E^{k,0} (V)$ represents the static total energy at 0 K without vibrational contribution, $F^{k,{vib}} (V,T)$ is the vibrational contribution to the Helmholtz energy, and $F^{k,{el}} (V,T)$ is the thermal electronic contribution to the Helmholtz energy \cite{Wang2004ThermodynamicCalculations, Shang2010First-principlesNi3Al}. Using \textbf{Eq. \ref{eq:1}}, various schemes of the quasiharmonic approximation (QHA) can be applied to calculate thermodynamic properties as a function of temperature and pressure, such as Gibbs energy, equilibrium volume, thermal expansion coefficient, entropy, specific heat capacity, enthalpy, and bulk modulus \cite{Shang2024Revisiting625}.

Multiple open-access databases, such as the Materials Project \cite{Jain2013Commentary:Innovation}, OQMD \cite{Kirklin2015TheEnergies, Saal2013MaterialsOQMD}, and AFLOW \cite{Curtarolo2012AFLOW:Discovery}, provide extensive datasets of material properties predicted through DFT calculations. However, these predictions are generally limited to high-symmetry configurations at equilibrium volume and 0 K temperature. For example, among the 153,902 materials available in the current Materials Project database, only 233 have energy-volume data fitted to equations of state (EOS) necessary to derive $E^{k,0}(V)$ \cite{Latimer2018EvaluationProject}, while the rest only have computed energies at their equilibrium volumes.
 
For certain materials, both high-symmetry and symmetry-breaking configurations are essential to describe material properties at finite temperatures accurately \cite{Liu2024ZentropyParameters}. These symmetry-breaking configurations, which include magnetic spin arrangements \cite{Shang2010ThermodynamicFe, Shang2010MagneticApproach, Wang2010ThermodynamicPrototype, Du2022DensityYNiO3} and dipole configurations in ferroelectric materials \cite{Hew2024PredictingCalculations, Zhao2022IntrinsicBaTiO3}, are metastable and can be accessed at finite temperatures. Such configurations are crucial for accurately capturing phase transitions without relying on fitted parameters \cite{Liu2024ZentropyParameters}.

We have previously developed the Density Functional Theory ToolKit (DFTTK) \cite{Wang2021DFTTK:Calculations} to accomplish two primary tasks to: (1) automate Helmholtz energy calculations using DFT with the widely-used Vienna Ab initio Simulation Package (VASP) \cite{Kresse1996EfficiencySet, Kresse1996EfficientSet}, and (2) store the Helmholtz energy of different configurations in a MongoDB database. The present paper presents a major revision of the previous DFTTK, focusing on updates to its core dependencies and the QHA scheme. These improvements were made to enhance DFTTK's user-friendliness and flexibility.

In this paper, we organize the sections as follows. Section 2 details the core software dependencies of DFTTK and the rationale behind changes from the previous version. Section 3 presents examples of running DFTTK for FCC \ch{Al} and magnetic \ch{Fe3Pt}, along with a breakdown of its key workflow components and details of the newly implemented QHA scheme. Section 4 concludes the paper with a summary and discussion of future work.

\section{Core dependencies}
\begin{figure}[ht]
    \centering
    \includegraphics[width=0.6\linewidth,trim=10 10 10 10,clip]{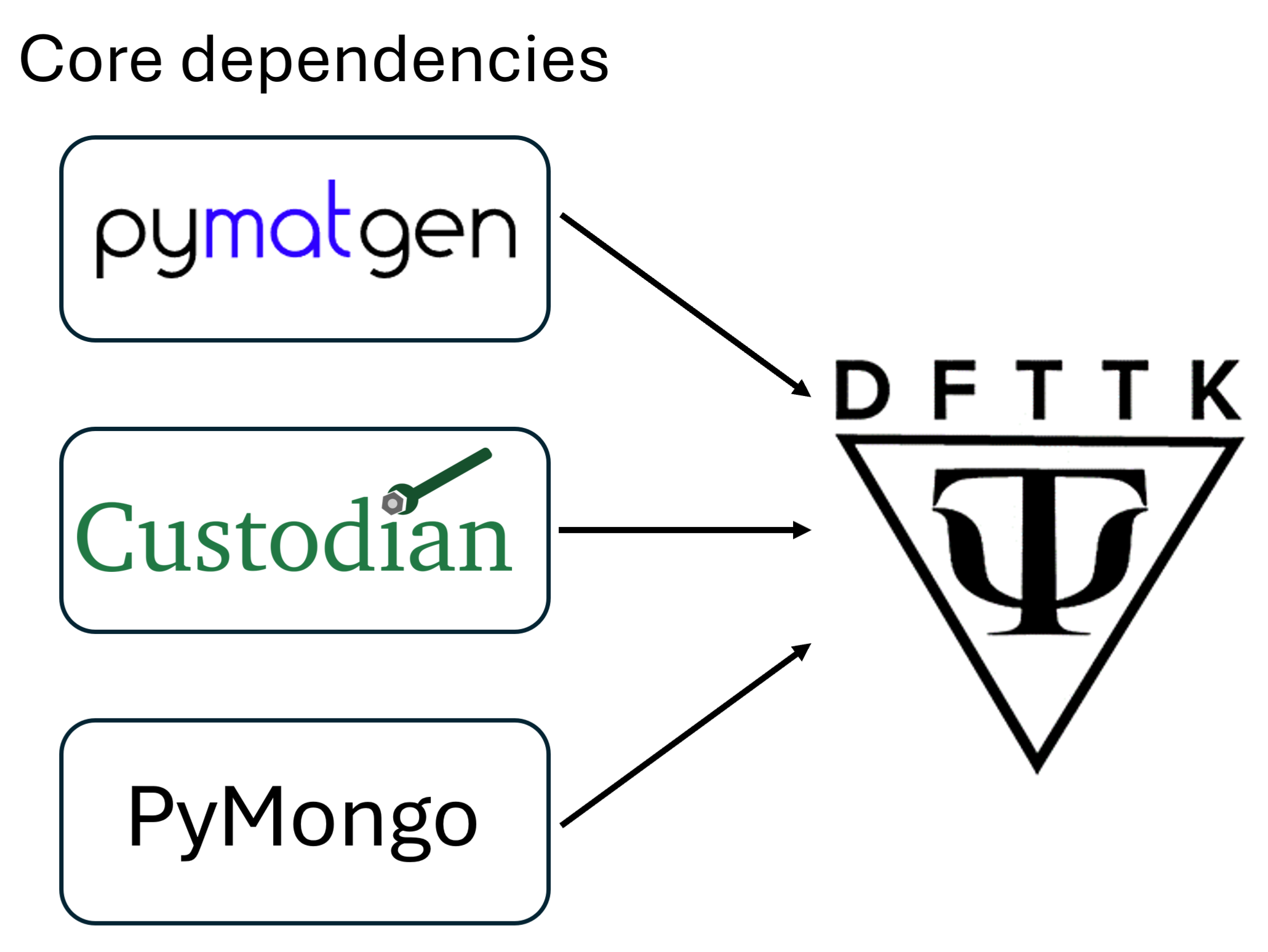}
    \caption{DFTTK independent core dependencies.}
    \label{DFTTK core dependencies}
\end{figure}

The previous version of DFTTK \cite{Wang2021DFTTK:Calculations} was primarily based on atomate \cite{Mathew2017Atomate:Workflows} workflows, which in turn relied on the Python packages pymatgen \cite{Ong2013PythonAnalysis} for handling VASP input/output and structure manipulation, Custodian \cite{Custodian} for running VASP jobs and error correction, and FireWorks \cite{Jain2015FireWorks:Applications} as a workflow manager to execute VASP jobs via Custodian and store results in a MongoDB database. In contrast, the current version of DFTTK no longer depends on atomate \cite{Mathew2017Atomate:Workflows}, though it still relies on pymatgen \cite{Ong2013PythonAnalysis} and Custodian \cite{Custodian}. Instead of FireWorks, PyMongo \cite{PyMongo} is now used to manage the storage of results in a MongoDB database. The independent core dependencies of the current DFTTK package are pymatgen\cite{Ong2013PythonAnalysis}, Custodian\cite{Custodian}, and PyMongo\cite{PyMongo}, as illustrated in \textbf{Figure \ref{DFTTK core dependencies}}.

The decision to remove the atomate \cite{Mathew2017Atomate:Workflows} and FireWorks \cite{Jain2015FireWorks:Applications} dependencies was made to streamline installation, usage, and debugging. With fewer installation steps, setup is now more straightforward. By eliminating FireWorks, users can run VASP jobs using Custodian workflows without needing a MongoDB connection. Previously, DFTTK required a MongoDB connection to run jobs, which caused issues if a cluster or supercomputer couldn't connect, preventing job execution. The new DFTTK adopts a modular "run first, store later" approach, allowing VASP jobs to run without MongoDB, with results stored later using PyMongo on any machine with MongoDB access. Of course, DFTTK and atomate can still be freely used in conjunction, for example, to generate data on a truly large scale. The current version of DFTTK simply removes the hard dependency between these packages.

This modular design simplifies debugging by isolating issues on the run and store sides. In addition, removing atomate \cite{Mathew2017Atomate:Workflows} and some of its dependencies has significantly simplified the codebase. Since FireWorks \cite{Jain2015FireWorks:Applications} itself relies on PyMongo \cite{PyMongo}, using PyMongo directly instead of FireWorks eliminates an additional layer of complexity, further streamlining the codebase.

\section{Running DFTTK}

\subsection{Installation and Access}  
DFTTK is publicly available on GitHub \cite{HewDFTTK}, where users can access the latest source code, report issues, and contribute to its development. A step-by-step installation guide is provided to facilitate setup. Users can install DFTTK using \texttt{pip install}. The repository also includes automated testing to ensure reliability. While this paper highlights some applications of DFTTK, additional tutorials and usage examples can be found in the official documentation \cite{HewDFTTKDocumentation}.  

\subsection{Workflow}
\begin{figure}[ht]
    \centering
    \includegraphics[width=1\textwidth]{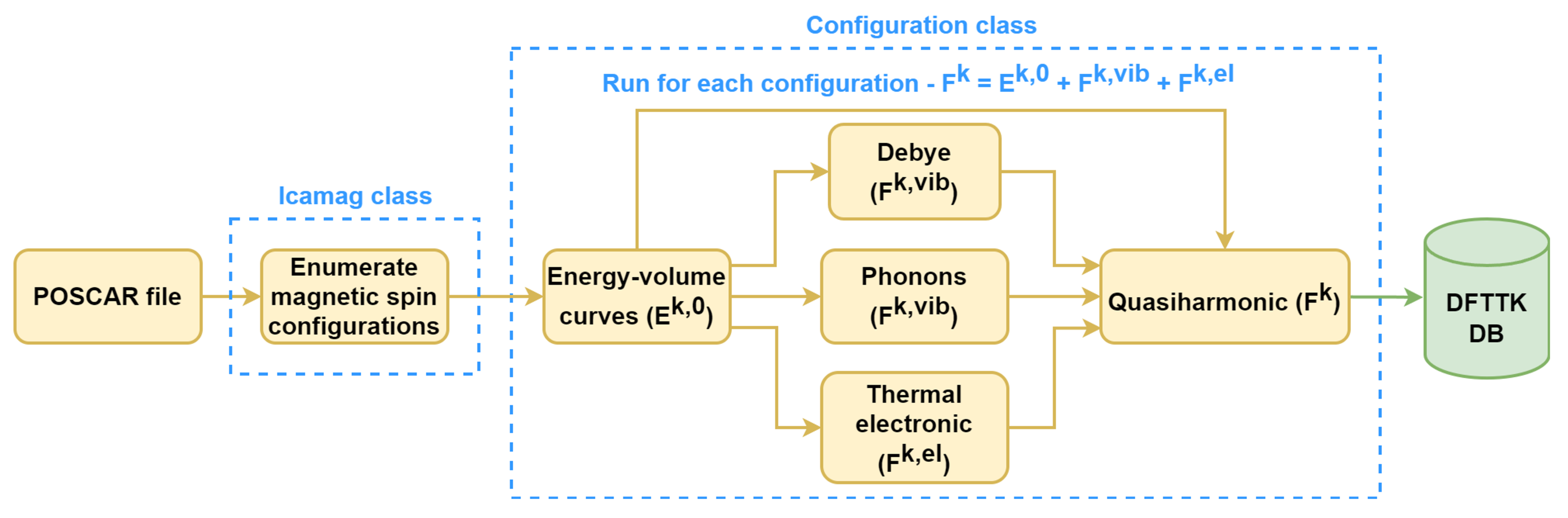}
    \caption{DFTTK Helmholtz energy workflow. A POSCAR file is used as input. For collinear magnetism, the \texttt{Icamag} class enumerates all unique magnetic spin configurations and their corresponding multiplicities. The \texttt{Configuration} class computes all three contributions to the Helmholtz energy: $E^{k,0}$, $F^{k, vib}$, and $F^{k,el}$, and applies the quasiharmonic approximation (QHA) to obtain $F^{k}$. Relevant results are then stored in the DFTTK MongoDB database.}
    \label{workflow}
\end{figure}

\textbf{Figure~\ref{workflow}} illustrates the Helmholtz energy workflow utilized in DFTTK to determine $F^{k}$ as described in \textbf{Eq.~\ref{eq:1}}. The only required input is a VASP-based structure file (POSCAR file) and arguments for methods. For magnetic structures, the \texttt{Icamag} (Independent Cell Approximation for MAGnetic systems) class, based on the "independent cells" approximation \cite{Pomrehn2011PredictedPhase}, generates all unique collinear magnetic spin configurations along with their respective multiplicities. 

The \texttt{Configuration} class then executes VASP jobs in series for each configuration to obtain the energy-volume data points, yielding $E^{k,0} (V)$. Subsequently, the Debye-Grüneisen model is applied to compute $F^{k,{vib}} (V,T)$. Phonon calculations can also be performed in parallel using VASP to more accurately evaluate $F^{k,{vib}} (V,T)$. Additionally, the thermal electronic contribution is obtained by running VASP jobs in parallel to calculate $F^{k,{el}} (V,T)$. Next, the QHA is applied to obtain the fitted $F^{k} (V,T)$, which is then used to calculate thermodynamic properties as functions of temperature and pressure. Details of the VASP input files and relevant properties are stored in the DFTTK MongoDB database. Note that each module in the \texttt{Configuration} class is designed to operate independently. For instance, the Debye module can be utilized with relevant inputs from other sources without needing to run the energy-volume curve calculations.

\textbf{Table \ref{tab:table_1}} presents some of the key methods in the \texttt{Icamag} and \texttt{Configuration} classes, along with their purposes. Further details and usage examples can be found on our GitHub page \cite{HewDFTTK}. The following subsections highlight the capabilities of each component in the workflow, using FCC Al with a 4-atom supercell and magnetic \ch{Fe3Pt} with a 12-atom 1x1x3 supercell as examples. For both materials, the Perdew-Burke-Ernzerhof (PBE) exchange-correlation functional \cite{Perdew1996GeneralizedSimple} was used. The version of the PBE pseudopotential files used by VASP (POTCAR files) employed was the version of PBE.54 for Al, $\text{Fe\textunderscore pv}$, and Pt with corresponding valence electrons of $3s^2 3p^1$, $3p^6 4s^2 3d^6$, and $5d^9 6s^1$, respectively. The electronic self-consistency loop ($\text{EDIFF}$) and the ionic relaxation loop ($\text{EDIFFG}$) convergence criteria were set to $1 \times 10^{-6}$ eV/supercell and $0.01$ eV/Å, respectively. The cutoff energy ($\text{ENCUT}$) was set to 520 eV. A k-point mesh of 4000 per reciprocal atom was used for Al, while 8800 per reciprocal atom was used for \ch{Fe3Pt}.

\subsection{Enumerate magnetic spin configurations}
The number of possible collinear magnetic configurations for a given supercell equals $2^n$, with $n$ being the number of magnetic atoms in the supercell. Take, for example, a 1x1x3 or 12-atom \ch{Fe3Pt} supercell with 9 Fe and 3 Pt atoms. Considering only the Fe as magnetic, the number of possible magnetic configurations is $2^9 = 512$. However, not all configurations are symmetrically unique. 

\begin{figure}[ht]
    \centering
    \includegraphics[width=0.6\textwidth]{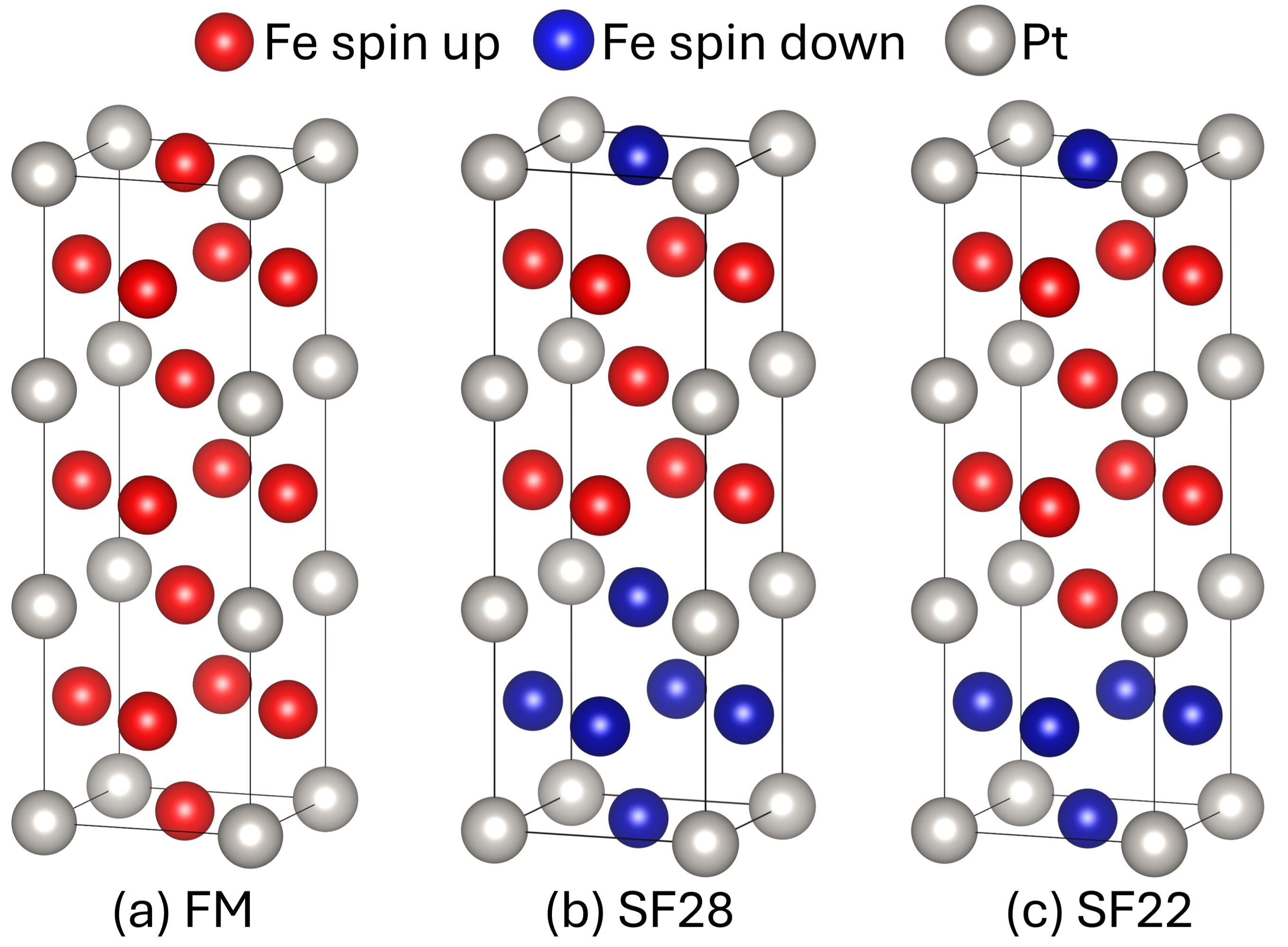}
    \caption{Three lowest energy $\text{Fe}_3\text{Pt}$ configurations using 12-atom supercells: (a) Ferromagnetic (FM) ground state, (b) spin-flipping (SF) 28, and (c) SF22. In the collinear case, spin-up and spin-down indicate the relative orientation of magnetization on each ion, with on-site magnetic moments having no defined direction in real space since spin-orbit coupling was not included.}
    \label{Fe3Pt configurations}
\end{figure}

For example, there are two ferromagnetic (FM) configurations: one where all spins are up and the other where all spins are down. These two configurations are symmetrically equivalent, meaning they can be considered as a single unique configuration. Consequently, the multiplicity of the all-spin-up configuration is 2. This unique FM configuration corresponds to the lowest energy or the ground state, as illustrated in \textbf{Figure~\ref{Fe3Pt configurations}a}. \textbf{Figure~\ref{Fe3Pt configurations}b} and \textbf{Figure~\ref{Fe3Pt configurations}c} depict the next two lowest energy configurations, labeled as SF28 and SF22, respectively. Here, SF denotes spin flipping, indicating that the number of spin-up magnetic moments differs from the number of spin-down magnetic moments, resulting in a net total magnetic moment. 

The \texttt{Icamag} class generates all 37 unique magnetic configurations for the \ch{Fe3Pt} system along with their associated multiplicities. Upon instantiating the class, these configurations and multiplicities are generated using the \texttt{gen\_spin\_configs(magnetic\_sites=\{"Fe": ["Fe+5", "Fe-5"]\})} and \texttt{get\_multiplicity()} methods, respectively. Calling \texttt{parse\_spin\_configs()} outputs key details of these configurations.

\begin{table}[ht]
    \centering
    \caption{Key methods in the \texttt{Icamag} and \texttt{Configuration} classes and their purposes.}  
    \label{tab:table_1}  
    \resizebox{\textwidth}{!}{%
    \begin{tabular}{l l l}  
        \toprule
        \textbf{Class} & \textbf{Method} & \textbf{Purpose} \\  
        \midrule
        \multirow{1}{*}{\texttt{Icamag}}  
        & \texttt{gen\_spin\_configs()} & Generates all unique collinear magnetic configurations. \\  
        & \texttt{parse\_spin\_configs()} & Outputs key details of all unique collinear magnetic configurations. \\  
        & \texttt{get\_multiplicity()} & Outputs the multiplicities of spin configurations. \\  
        \midrule
        \multirow{1}{*}{\texttt{Configuration}}  
        & \texttt{ev\_curve\_settings()} & Configures settings for EV curve calculations. \\  
        & \texttt{run\_ev\_curve()} & Runs EV curve calculations. \\  
        & \texttt{process\_ev\_curve()} & Processes and extracts data from EV calculations. \\ 
        & \texttt{ev\_curve.eos\_parameters()} & Outputs the EOS parameters. \\ 
        & \texttt{ev\_curve.plot()} & Plots EV curve. \\
        \cmidrule{2-3}
        & \texttt{process\_debye()} & Processes Debye-Grüneisen results. \\
        & \texttt{debye.plot()} & Plots Debye-Grüneisen results. \\
        \cmidrule{2-3}
        & \texttt{phonons\_settings()} & Configures settings for phonon calculations. \\  
        & \texttt{run\_phonons()} & Runs phonon calculations. \\  
        & \texttt{generate\_phonon\_dos()} & Computes the phonon density of states (DOS). \\  
        & \texttt{process\_phonons()} & Processes and extracts data from phonon calculations. \\  
        & \texttt{phonons.plot\_harmonic()} & Plots harmonic phonon results. \\
        \cmidrule{2-3}
        & \texttt{thermal\_electronic\_settings()} & Configures settings for thermal electronic calculations. \\  
        & \texttt{run\_thermal\_electronic()} & Runs thermal electronic calculations. \\  
        & \texttt{process\_thermal\_electronic()} & Processes thermal electronic results. \\  
        & \texttt{thermal\_electronic.plot()} & Plots thermal electronic results. \\  
        \cmidrule{2-3}
        & \texttt{process\_qha()} & Processes quasiharmonic approximation (QHA) results. \\  
        & \texttt{qha.plot()} & Plots QHA results. \\  
        \cmidrule{2-3}
        & \texttt{to\_mongodb()} & Exports input and processed data to a MongoDB database. \\ 
        \bottomrule
    \end{tabular}%
    }
\end{table}

\subsection{Energy-volume curves}
The first term $E^{k,0} (V)$ in \textbf{Eq. \ref{eq:1}} is calculated from an EOS fitted to DFT data points. A commonly used EOS is the four-parameter Birch-Murnaghan (BM4) EOS \cite{Shang2010First-principlesNi3Al}:
\begin{equation}
    E^{k,0} (V) = a + bV^{-2/3} + cV^{-4/3} + dV^{-2},
\label{eq:2}
\end{equation}

where $a$, $b$, $c$, and $d$ are the fitting parameters. Knowing this, the corresponding equilibrium volume, $V_0$, bulk modulus, $B_0$, and the derivative of the bulk modulus with respect to pressure, $ B'_0$, can be determined:

\begin{equation}
    V_0 = \sqrt{\frac{9bcd - 4c^3 - \sqrt{\left(c^2-3bd\right)\left(4c^2-3bd\right)^2}}{b^3}}
\label{eq:3}
\end{equation}

\begin{equation}
    B_0 = \frac{2\left(27d + 14cV_0^{2/3} + 5bV_0^{4/3}\right)}{9V_0^3}
\label{eq:4}
\end{equation}

\begin{equation}
    B_0^{'} = \frac{243d + 98cV_0^{2/3} + 25bV_0^{4/3}}{81d + 42cV_0^{2/3} + 15bV_0^{4/3}}
\label{eq:5}
\end{equation}

The equilibrium energy, $E_0$, can be calculated by substituting $V_0$ into \textbf{Eq. \ref{eq:2}}. 

\begin{figure}[ht]
    \centering
    \includegraphics[width=1\textwidth]{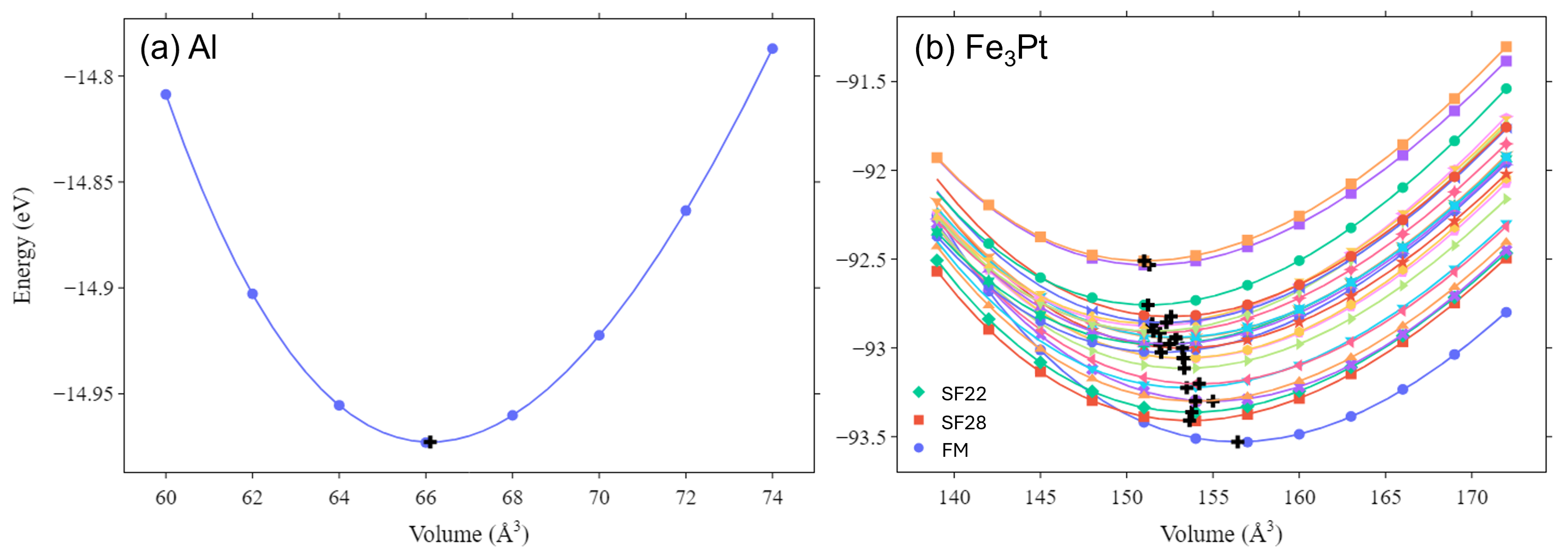}
    \caption{Energy-volume curves for (a) Al using a 4-atom supercell and (b) $\text{Fe}_3\text{Pt}$ magnetic configurations using 12-atom supercells. The solid points represent the data from DFT, while the curves are from the fitted BM4 EOS to those points. The $\fplus$ symbols indicate the energy-volume curve minima. Only the three lowest energy configurations for $\text{Fe}_3\text{Pt}$ are labeled for brevity.}
    \label{ev curves}
\end{figure}

\textbf{Figure~\ref{ev curves}} displays the energy-volume curves for Al and \ch{Fe3Pt}, 
with solid points representing DFT results and the curves corresponding to the fitted BM4 EOS. 
Upon instantiating the \texttt{Configuration} class, the calculation settings can be defined 
using the \texttt{ev\_curve\_settings(material\_type="metal", volumes, encut,} 
\texttt{kppa, other\_settings, copy\_magmom=True)} method and executed with \texttt{run\_ev\_curve()}. 
After the calculations are complete, the results can be processed using 
\texttt{process\_ev\_curve(collect\_mag\_data=True)}, where \texttt{collect\_mag\_data} is set 
to \texttt{True} if you want to collect magnetic data for magnetic calculations. The EOS for a 
single configuration can be visualized with \texttt{ev\_curve.plot()}, and for multiple configurations, 
the \texttt{plot\_multiple\_ev(config\_objects,\allowbreak config\_names)} function can be used 
to generate comparative plots.

DFTTK employs a sequential approach, starting with relaxation at one volume and using the relaxed structure from that volume as the initial structure for the next volume. We recommend, particularly for magnetic materials, beginning with the highest volume of study first to prevent relaxation into a different magnetic configuration, as magnetic moments tend to be stabilized at higher volumes. If the opposite behavior is observed, one can begin with the lowest volume instead. However, this transition is unavoidable for certain configurations, causing them to jump to another configuration between the volume steps of the energy-volume curve calculations. For instance, in \textbf{Figure~\ref{ev curves}b}, 12 configurations were excluded because they relaxed into a different configuration before reaching their equilibrium volumes, which hindered obtaining a good EOS fit. Nonetheless, these were higher energy configurations and are thus less significant. For clarity, only the three lowest energy configurations are labeled in \textbf{Figure~\ref{ev curves}b}.

The equilibrium properties were calculated using the BM4 EOS, accessible through 
\texttt{ev\_curve.eos 
\_parameters()}, for Al and the three lowest energy configurations of \ch{Fe3Pt} are provided in \textbf{Table~\ref{tab:BM4_EOS_parameters}}. The BM4 EOS parameters for all stable \ch{Fe3Pt} configurations are shown in \textbf{Table~\ref{tab:Fe3Pt_eos}} in \textbf{Appendix A}. In addition, DFTTK supports various other equations of state, including the four- and five-parameter Teter-Shang modified Birch-Murnaghan (mBM4 and mBM5), five-parameter Birch-Murnaghan (BM5), four- and five-parameter logarithmic (LOG4 and LOG5), Murnaghan, Vinet, and Morse EOS, as described by Shang et al. \cite{Shang2010First-principlesNi3Al}.

\begin{table}[ht]
\centering
\caption{BM4 EOS parameters $\mathrm{E_0}$, $\mathrm{V_0}$, $\mathrm{B_0}$, and $\mathrm{B'_0}$ for Al and for the three lowest energy configurations of $\text{Fe}_3\text{Pt}$. $\mathrm{\Delta E}$ refers to the energy difference using the ground state as the reference.}
\label{tab:BM4_EOS_parameters}
\small
\begin{adjustbox}{max width=\textwidth}
\begin{tabular}{ccccccc}
\toprule
Material & Configuration & $\mathrm{E_0}$ (eV/atom) & $\mathrm{\Delta E}$ (meV/atom) & $\mathrm{V_0}$ (\AA$^3$/atom) & $\mathrm{B_0}$ (GPa) & $\mathrm{B'_0}$ \\ 
\midrule
\multirow{1}{*}{\textbf{\ch{Al}}}
& FCC & -3.7432 & 0 & 16.5255 & 77.9279 & 4.6127 \\ 
\midrule
\multirow{3}{*}{\textbf{\ch{Fe3Pt}}}
& FM   & -7.7941 & 0       & 13.03605 & 175.5191 & 3.6599 \\ 
& SF28 & -7.7841 & 9.9771  & 12.8034  & 162.7900 & 4.1881 \\ 
& SF22 & -7.7801 & 13.9109 & 12.8142  & 161.8699 & 4.2696 \\ 
\bottomrule
\end{tabular}
\end{adjustbox}
\end{table}

\subsection{Vibrational contribution by Debye-Grüneisen model}
The second term $F^{k,{vib}} (V, T)$ in \textbf{Eq. \ref{eq:1}} can be calculated via the empirical Debye-Grüneisen model from the EOS parameters. According to this model, the Helmholtz energy is given by \cite{Shang2010First-principlesNi3Al}

\begin{equation}
    F^{k, {vib}} (V, T) = \frac{9}{8}k_B\Theta_D + k_BT \left[ 3 \ln \left( 1 - \exp \left( - \frac{\Theta_D}{T} \right) \right) - D \left( \frac{\Theta_D}{T} \right) \right],
\label{eq:10}
\end{equation}
where $\Theta_D$ is the Debye temperature given by  
\begin{equation}
    \Theta_D = sAV_0^{1/6} \left( \frac{B_0}{M} \right) ^{1/2} \left( \frac {V_0}{V}\right) ^\gamma,
\label{eq:11}
\end{equation}

where $s$ is a scaling factor. The default value for $s$ is 0.617 obtained for non-magnetic cubic metals \cite{Moruzzi1988CalculatedMetals}. Alternatively, the scaling factor, $s$, can be estimated from phonon calculations \cite{Shang2010First-principlesNi3Al} or from elastic constants \cite{Liu2015OnSystem}. The terms $V_0$ and $B_0$ are equilibrium volume and bulk modulus obtained from the EOS fit. $M$ is the average atomic mass and $\gamma$ is the Grüneisen constant given by 

\begin{equation}
    \gamma = \frac{1 + B_0'}{2} - x,
\label{eq:12}
\end{equation}

where $x = 2/3$ is typically used for higher temperatures and $x = 1$ for lower temperatures \cite{Moruzzi1988CalculatedMetals}. The term $ B'_0$ is shown in \textbf{Eq. \ref{eq:5}} from the EOS fit. The parameter $A$ is a constant given by 

\begin{equation}
    A = \frac{(6\pi^2)^{1/3} \hbar}{k_B},
\label{eq:13}
\end{equation}

where $\hbar$ is the reduced Planck's constant. The last parameter that needs to be defined is $D (\Theta_D/T)$, which is the Debye function:

\begin{equation}
    D(\Theta_D/T) = \frac{3}{(\Theta_D/T)^3} \int_{0}^{\Theta_D/T} \frac{t^3}{\exp(t)-1} \, dt.
\label{eq:14}
\end{equation}

Given $F^{k, \text{vib}}(V, T)$, other thermodynamic properties can be calculated, such as the vibrational entropy $S^{k, \text{vib}}(V, T)$ and the vibrational specific heat capacity at constant volume $C_v^{k, \text{vib}}(V, T)$. Calling \texttt{process\_debye(scaling\_factor, gruneisen\_x, temperatures)} applies the Debye--Grüneisen model to compute these properties, which can then be visualized using \texttt{debye.plot()}.

\subsection{Vibrational Contribution by Phonons}
Alternatively, the term $F^{k,{vib}} (V, T)$ can also be calculated using phonons. If one has the phonon density of states (DOS) calculated from DFT, $F^{k,{vib}} (V, T)$ can be calculated from Bose-Einstein statistics \cite{Shang2010First-principlesNi3Al}:

\begin{equation}
    F^{k,{vib}} (V, T) = k_B T \int_{0}^{\infty} \ln \left[ 2 \sinh \left( \frac{h v}{2 k_B T} \right) \right] g(v) \, dv,
\label{eq:7}
\end{equation}

where $k_B$ is the Boltzmann constant, $h$ is Planck's constant, and $g(v)$ is the phonon DOS as a function of the phonon frequency, $v$. The vibrational entropy, $S^{k, \text{vib}}(V, T)$, and the vibrational heat capacity at constant volume, $C_v^{k, \text{vib}}(V, T)$, can also be calculated if $g(v)$ is known. The phonon settings are configured using \texttt{phonons\_settings(phonon\_volumes, kppa, scaling\_matrix)} and run in parallel with \texttt{run\_phonons()}. Currently, DFTTK uses the YPHON package to calculate the phonon DOS after VASP calculations \cite{YPHON, Wang2014YPHON:Materials} using \texttt{generate\_phonon\_dos()}.

\begin{figure}[ht]
    \centering
    \includegraphics[width=1\textwidth]{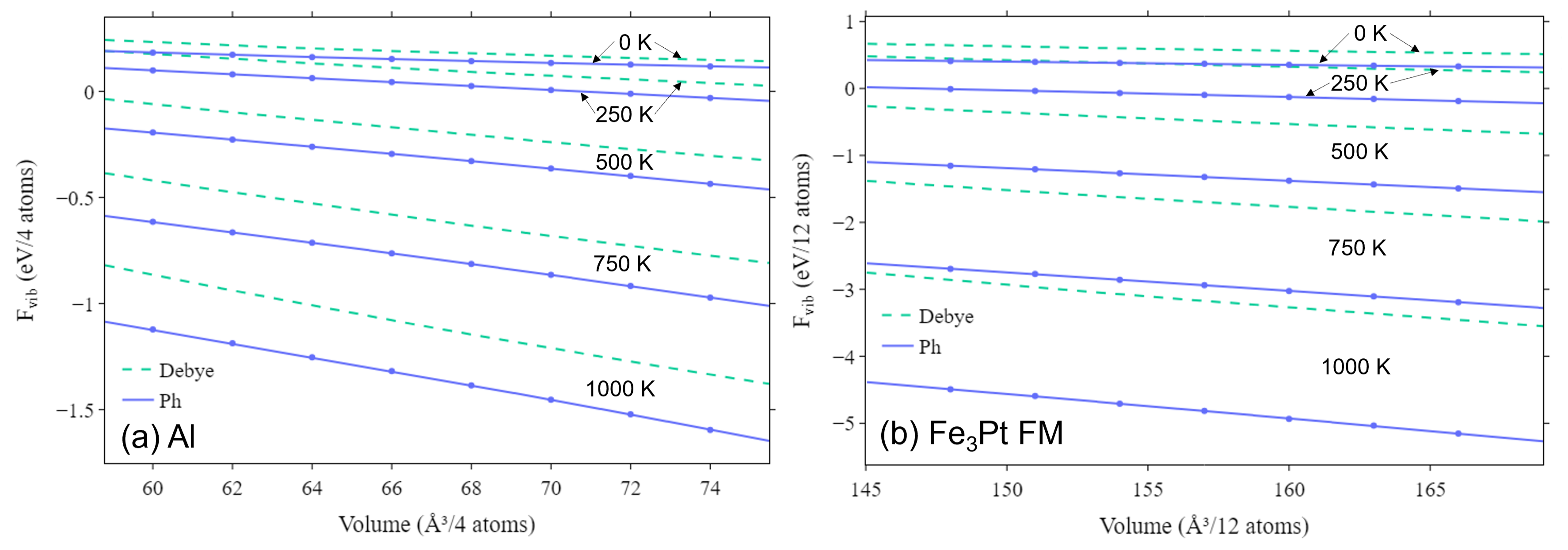}
    \caption{Vibrational contribution to the Helmholtz energy using the Debye–Grüneisen model and phonons calculations for (a) Al using a 4-atom supercell and (b) $\text{Fe}_3\text{Pt}$ ferromagnetic (FM) ground-state a 12-atom supercell. Values of $s=0.617$ and $x=2/3$ were used for the Debye–Grüneisen model.}
    \label{fvib}
\end{figure}

\textbf{Figure~\ref{fvib}} presents the $F^{k, \text{vib}}$ results for a few fixed 
temperature curves for Al (\textbf{Figure~\ref{fvib}a}) and the FM ground state of 
$\text{Fe}_3\text{Pt}$ (\textbf{Figure~\ref{fvib}b}) calculated using both the 
Debye–Grüneisen model ($s=0.617$ and $x=2/3$) and phonon calculations. The phonon 
results are processed using 
\texttt{process\_phonons(number\_of\_atoms,\allowbreak temperatures)} 
and plotted using 
\texttt{phonons.\allowbreak plot\_harmonic()}. As expected, the Debye–Grüneisen model 
agrees reasonably well with the phonon results at lower temperatures, with deviations 
increasing as the temperature rises~\cite{Shang2010First-principlesNi3Al}. In DFTTK, a 
polynomial function is fitted to $F^{k, \text{vib}}$ as a function of volume for each 
fixed-temperature curve, with the fitting order determined by the user. Future improvements 
could incorporate the corrected Akaike information criterion (AICc)~\cite{Cavanaugh1997UnifyingCriteria} 
to optimize the polynomial order selection and mitigate overfitting, similar to the approach 
used in the Python package ESPEI~\cite{Bocklund2019ESPEICuMg} and our previous efforts~\cite{Shang2024Revisiting625}.

\subsection{Thermal electronic}
The third term in \textbf{Eq.~\ref{eq:1}} is the thermal electronic contribution to the Helmholtz energy, $F^{k,{el}} (V, T)$, which can be determined by Mermin statistics \cite{Shang2010First-principlesNi3Al},

\begin{equation}
    F^{k,el} (V,T) = E^{k,el} - TS^{k,el}.
\label{eq:17}
\end{equation}

The term $E^{k,el}$ is the internal energy due to electronic excitation for each configuration $k$:

\begin{equation}
    E^{k,el} (V,T) = \int_{-\infty}^{\infty} n\left(\epsilon\right) f
    \epsilon\, d\epsilon - \int_{-\infty}^{\epsilon_F} n\left(\epsilon\right) \epsilon\, d\epsilon,
\label{eq:18}
\end{equation}

where $n\left(\epsilon\right)$ is the electronic DOS, $\epsilon$ the energy eigenvalues, and $\epsilon_F$ is the Fermi energy. The function $f\left(\epsilon, T, V)\right)$ is called the Fermi distribution function given by 

\begin{equation}
    f\left(\epsilon, T, V)\right) = 1/ \left[\exp\left(\frac{\epsilon-\mu\left(V, T\right)}{k_BT}\right) + 1 \right],
\label{eq:19}
\end{equation}

where $\mu (V, T)$ is the electronic chemical potential. Lastly, the thermal electronic contribution to the entropy is written as 

\begin{equation}
    S^{k,el} (V, T) = -k_B \int_{-\infty}^{\infty} n\left(\epsilon\right) \left[f \ln{f} + \left(1-f\right) \ln{(1-f)}\right]\, d\epsilon.
\label{eq:20}
\end{equation}

For metals, the term $F^{k, \text{el}}(V, T)$ contributes only a small fraction to the total Helmholtz energy of a configuration, $F^{k}(V, T)$, in \textbf{Eq.~\ref{eq:1}} and is negligible for semiconductors and insulators. The thermal electronic settings are configured using \texttt{thermal\_electronic\_settings(volumes, kppa, scaling\_matrix)} and run in parallel using \texttt{run\_thermal\_electronic()}.

\begin{figure}[ht]
    \centering
    \includegraphics[width=1\textwidth]{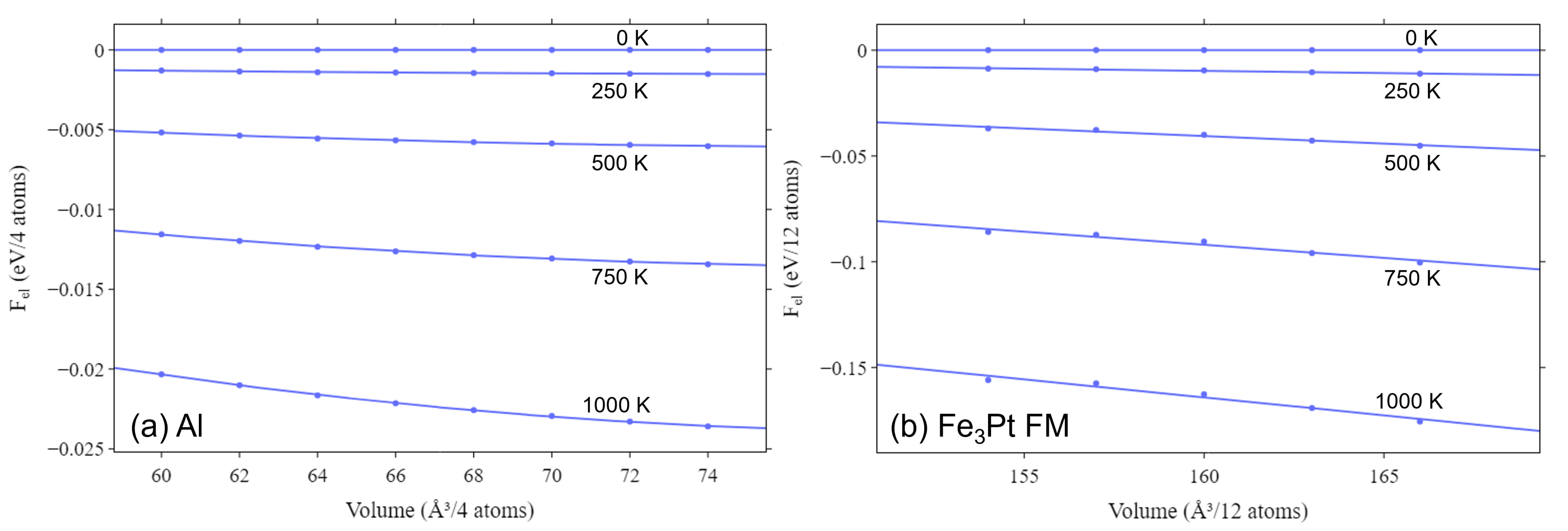}
    \caption{Thermal electronic contribution to the Helmholtz energy for (a) Al using a 4-atom supercell and (b) $\text{Fe}_3\text{Pt}$ ferromagnetic (FM) ground-state a 12-atom supercell.}
    \label{fel}
\end{figure}

\textbf{Figure~\ref{fel}} shows $F^{k,{el}}$ for several fixed-temperature curves for \ch{Al} (\textbf{Figure~\ref{fel}a}) and the FM ground state of \ch{Fe_3Pt} (\textbf{Figure~\ref{fel}b}). 
The thermal electronic
results are processed using \texttt{process\_thermal\_electronic(temperatures)} and plotted using \texttt{thermal\_electronic.plot()}.
Even in these metallic systems, the contribution of $F^{k,{el}}$ to the total $F^k$ is between 1 and 2 orders of magnitude smaller than that of $F^{k,{vib}}$.This demonstrates that the vibrational component, $F^{k,{vib}}$, plays a significantly more dominant role in determining $F^{k} (V, T)$, particularly at lower temperatures, where lattice vibrations are more prominent. Thus, the dominant contributions to $F^{k} (V, T)$ is $E^{k,0}(V)$, followed by $F^{k,{vib}}$ and $F^{k,{el}}$. This is evident when comparing \textbf{Figure~\ref{ev curves}b}, \textbf{Figure~\ref{fvib}b}, and \textbf{Figure~\ref{fel}b}. 

\subsection{Quasiharmonic approximation}
With all three contributions to the Helmholtz energy available, $F^k (V, T)$ can be calculated using \textbf{Eq. \ref{eq:1}}. In DFTTK, the fitted values from each contribution are combined as suggested by Shang et al. \cite{Shang2024Revisiting625}, rather than using the raw data points directly, as is done in other packages such as Phonopy \cite{Togo2023First-principlesPhono3py}. This approach offers several advantages \cite{Shang2024Revisiting625}, including generating data points for energy-volume curves, phonons, and electronic density of states (DOS) at different volumes and/or with fewer volume points. This is particularly beneficial for phonon calculations, which are the most computationally expensive of the three. By using fitted values, interpolation and extrapolation can be performed with fewer phonon data points while still capturing the overall trends, thereby reducing computational costs without compromising essential behavior.

To calculate the Gibbs energy at a given pressure, $P$, the term $PV$ for a certain volume range is first added to $F^k(V, T)$, yielding $F^k + PV$. An EOS is then used to fit each isothermal $F^k(V, T)$ curve, determining the equilibrium volume, $V^{k}_{c}(T, P)$, and the corresponding minimum energy, or Gibbs energy, $G^{k}(T, P)$, at that fixed pressure, as expressed in the equations below \cite{Togo2023First-principlesPhono3py}:
\begin{equation}
    G^k (T, P) = \min_{V} \left[ F^k + PV \right]
\label{eq:21}
\end{equation}
\begin{equation}
    V^k_c (T, P) = \arg \min_{V} \left[ F^k + PV \right].
\label{eq:22}
\end{equation}

From \textbf{Eq. \ref{eq:22}}, the volumetric coefficient of thermal expansion (CTE), $\alpha_V^{k} (T, P)$, can be obtained as:
\begin{equation} \alpha_V^{k}(T, P) = \frac{1}{V^k_c} \frac{dV^k_c}{dT} \label{eq:23} \end{equation}

Given the total entropy of a configuration corresponding to $V^k_c$, where $S^k = S^{k, vib} + S^{k, el}$, the specific heat capacity at constant pressure, $C_p^{k} (T, P)$, can be calculated using the first equality in the following equation:
\begin{equation} C_p^{k} (T, P) = T \frac{dS^k}{dT} = C_v^{k} + (\alpha_V^{k})^2 B_c^{k} T V_c^{k} \label{eq:24} \end{equation}

Alternatively, $C_p^{k} (T, P)$ can also be calculated using the second equality in \textbf{Eq. \ref{eq:24}}, where $C_v^{k}$, the specific heat capacity at constant volume, and $B_c^{k}$, the bulk modulus, are both evaluated at the equilibrium volume, $V_c^{k}$, at pressure, $P$. The enthalpy, $H^{k} (T, P)$, can then be obtained using the following equation:
\begin{equation}
    H^{k} (T, P) = G^{k} + TS^{k}
\label{eq:25}
\end{equation}

\begin{figure}[ht]
    \centering
    \includegraphics[width=1\textwidth]{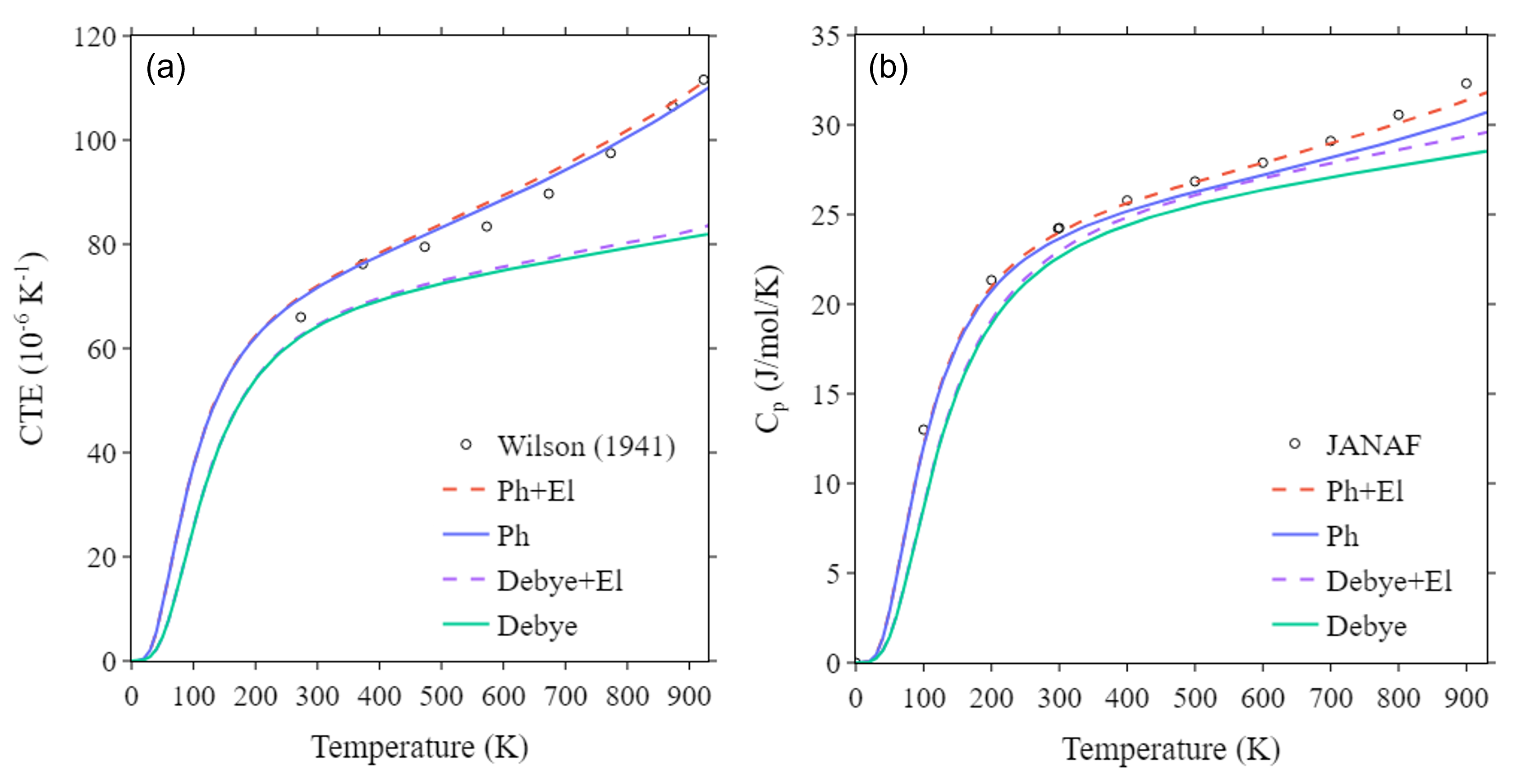}
    \caption{(a) Coefficient of thermal expansion (CTE) and (b) specific heat capacity at zero pressure, $C_p$, of fcc Al, based on experimental data from Wilson \cite{Wilson1941TheC} and the JANAF thermochemical tables \cite{NIST-JANAFThermochemicalTables1983AluminumAl}, as well as calculations from this work using the Debye-Grüneisen model (s=0.617, x=2/3), phonons, and thermal electronic contributions.
}
    \label{Al_expt}
\end{figure}

\textbf{Figure~\ref{Al_expt}} presents the results for CTE (\textbf{Figure~\ref{Al_expt}a}) and $C_p$ (\textbf{Figure~\ref{Al_expt}b}) of Al at zero external pressure, obtained using the Debye–Grüneisen model ($s=0.617$, $x=2/3$), phonons, and thermal electronic contributions, in comparison with experimental data from Wilson \cite{Wilson1941TheC} and the JANAF thermochemical tables \cite{NIST-JANAFThermochemicalTables1983AluminumAl}. The QHA results are processed using \texttt{process\_qha(method, volume\_range, P)} and plotted using \texttt{qha.plot()}.

It can be observed that the Debye-Grüneisen model underpredicts the CTE. This effect is related to \textbf{Figure~\ref{fvib}a}, as the slower decrease of $F^k_{vib}$ with increasing temperature leads to a smaller increase in equilibrium volume, thereby resulting in a lower CTE as a function of temperature. Additionally, the inclusion of the $F^k_{el}$ term only slightly increases the CTE for both the Debye-Grüneisen model and phonons, as its contribution to the overall $F^k$ remains relatively small. Similarly, the Debye-Grüneisen model underpredicts $C_p$ compared to phonons. The inclusion of the $F^k_{el}$ term increases $C_p$ for both the Debye-Grüneisen model and phonons, and this addition to the phonons enhances the agreement with experimental data at higher temperatures.

\begin{figure}[ht]
    \centering
    \includegraphics[width=0.6\textwidth]{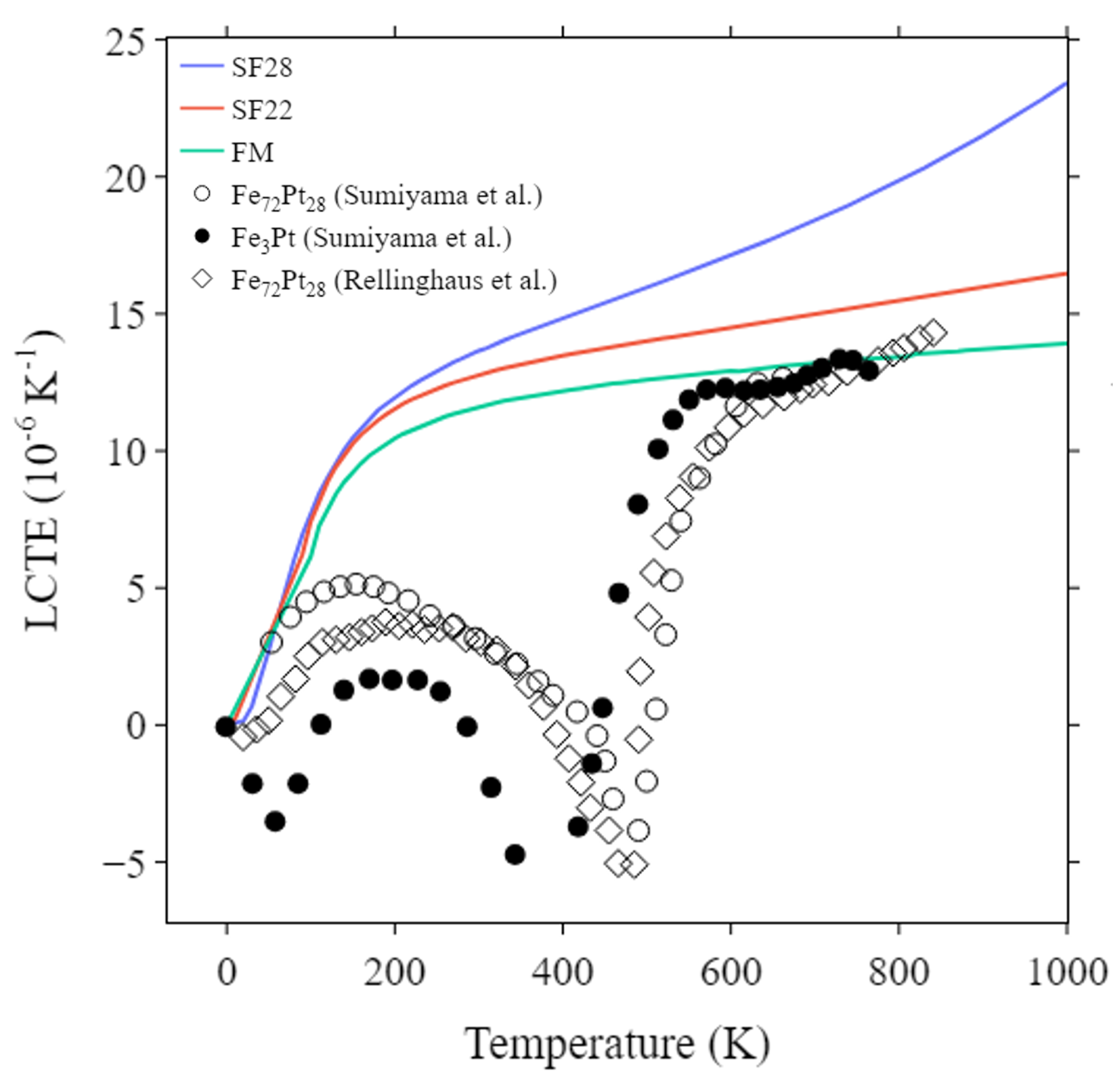}
    \caption{Linear coefficient of thermal expansion (LCTE) predicted in this work using phonons and the thermal electronic contribution for the ferromagnetic (FM) ground state, as well as for the next two lowest-energy spin-flipping (SF) configurations, SF22 and SF28. Experimental data for $\text{Fe}_{72}\text{Pt}_{28}$ from Sumiyama et al. \cite{Sumiyama1981MagneticBoundary} and Rellinghaus \cite{Rellinghaus1995ThermodynamicInvar}, as well as for $\text{Fe}_3\text{Pt}$ from Sumiyama et al. \cite{Sumiyama1979CharacteristicAlloys}, are also provided for comparison.}
    \label{Fe3Pt_expt}
\end{figure}

The linear coefficient of thermal expansion (LCTE) values (CTE/3) for the FM, SF28, and SF22 configurations of $\text{Fe}_3\text{Pt}$, computed using phonons and thermal electronic contribution, along with experimental data for $\text{Fe}_{72}\text{Pt}_{28}$ and $\text{Fe}_3\text{Pt}$ \cite{Sumiyama1981MagneticBoundary, Sumiyama1979CharacteristicAlloys, Rellinghaus1995ThermodynamicInvar}, are presented in \textbf{Figure~\ref{Fe3Pt_expt}}. The figure indicates that all three configurations computed in this work exhibit positive thermal expansions. This is in contrast to the experimental results, which show regions of both positive and negative thermal expansion, with a range sufficiently low for $\text{Fe}_3\text{Pt}$ to be considered as an Invar alloy. The Invar effect refers to the unusually low thermal expansion observed in these alloys over a wide temperature range \cite{Sumiyama1981MagneticBoundary, Sumiyama1979CharacteristicAlloys, Rellinghaus1995ThermodynamicInvar}.

\vspace{1em}
This apparent discrepancy underscores a limitation of the common DFT approach of using a single configuration to model the properties of a real material, which, in practice, consists of a mixture of configurations \cite{Liu2024ZentropyParameters, Liu2022ZentropyExpansion}. It can be demonstrated that the experimental Invar behavior is replicated by considering a statistical mixture of all possible \ch{Fe3Pt} configurations, each weighted by its respective probability \cite{Wang2010ThermodynamicPrototype, Shang2023QuantifyingFe3Pt}. As shown in \textbf{Figure~\ref{ev curves}b}, all non-ground state metastable configurations have equilibrium volumes smaller than the FM ground state. The regions of negative thermal expansion are explained by the increasing probabilities of these metastable configurations at higher temperatures, as the system's volume is a probability-weighted sum of the volumes of the individual configurations. This methodology, known as the zentropy approach \cite{Liu2024ZentropyParameters, Liu2022ZentropyExpansion}, calculates a system's entropy by taking the weighted sum of the contributions from the different configurations based on their respective probabilities and includes the computation of the configurational entropy. A new open-source package, PyZentropy, is currently being developed to implement this approach.

\subsection{MongoDB}
A crucial aspect of scientific communication is data accessibility, which ensures further analysis and reproducibility by other researchers. In computational research, this typically involves sharing calculated results, raw data, computer code, and any additional necessary information used to generate results \cite{Stodden2018AnReproducibility.}. For reproducing results from DFTTK, this includes the VASP input files, as well as the input parameters and workflows used in DFTTK.

To achieve this, we store the necessary results in our DFTTK MongoDB database using 
\texttt{to\_mongodb(connection\_string,\allowbreak db\_name,\allowbreak collection\_name)}. 
This database serves as a comprehensive repository, containing information on all VASP input
files used to generate results, along with fitted EOS parameters. Additionally, it includes 
thermodynamic properties derived from the Debye–Grüneisen model, phonon calculations, thermal 
electronic contributions, and quasiharmonic approximation results. We anticipate that the 
accumulation of this data will be of great interest for use in machine learning models and 
materials discovery \cite{Schleder2019FromReview}. Maintaining this structured database 
facilitates data accessibility, reproducibility, and further analysis by the research community.

Our current database contains 1753 energy-volume curve calculations and corresponding EOS parameters, and this number continues to grow. Our goal is to expand this repository to include phonon calculations, thermal electronic contributions, and quasiharmonic approximations, ultimately building a comprehensive Helmholtz energy database of configurations. This effort parallels our development of the Material-Property-Descriptor Database (MPDD) \cite{Material-Property-DescriptorMPDD}, which currently covers approximately 4.5 million DFT-relaxed structures and formation energies.

\section{Conclusions}
DFTTK was developed to automate Helmholtz energy calculations of individual configurations using first-principles calculations via VASP. For collinear magnetic calculations, it can generate all unique magnetic configurations and their associated multiplicities using the \texttt{Icamag} class, as demonstrated in the \ch{Fe_3Pt} example. The \texttt{Configuration} class enables the computation of all three contributions to the Helmholtz energy through energy-volume curves, the Debye-Grüneisen model, phonons, and thermal electronic contributions. Once these contributions are determined, the quasiharmonic approximation is applied, and the Helmholtz energy as a function of volume and temperature is stored in the DFTTK MongoDB database.

DFTTK is developed as an open-source package on GitHub. Its modular design allows for the seamless addition of new workflows, making it adaptable for further development. The Helmholtz energy of different configurations can also serve as input to the zentropy approach, which will be implemented in our upcoming PyZentropy package, currently under development.

\section*{Acknowledgments}
This work was supported by the Department of Energy (DOE) with Award Nos. DE-AR0001435 and DE-NE0009288, and by the National Science Foundation with Award Nos. CMMI-2226976, CMMI-2050069, and FAIN-2229690. Computational resources were provided by the Bridges-2 supercomputer at the Pittsburgh Supercomputing Center and the Stampede3 supercomputer at the Texas Advanced Computing Center through project DMR140063: Fundamental Properties of Crystalline Materials by First Principles from the Advanced Cyberinfrastructure Coordination Ecosystem: Services \& Support (ACCESS) program~\cite{10.1145/3569951.3597559}, supported by National Science Foundation grants 2138259, 2138286, 2138307, 2137603, and 2138296. Additional computations were performed on the Roar Collab supercomputer at the Pennsylvania State University's Institute for Computational and Data Sciences (ICDS). The content is solely the responsibility of the authors and does not necessarily reflect the views of ICDS~\cite{2025InstituteSciences}.

\section*{CRediT author statement}
\textbf{Nigel Lee En Hew:} Conceptualization, Methodology, Software, Validation, Formal analysis, Data Curation, Writing - Original Draft, Writing - Review \& Editing, Visualization. \textbf{Luke Allen Myers:} Conceptualization, Methodology, Software, Validation, Formal analysis, Writing - Review \& Editing \textbf {Axel van de Walle:} Software - developed \texttt{Icamag}. \textbf{Shun-Li Shang:} Conceptualization, Writing - Review \& Editing, Supervision. \textbf{Zi-Kui Liu:} Conceptualization, Resources, Supervision, Writing - Review \& Editing, Funding acquisition.

\printbibliography \label{bib}

\newpage
\appendix
\renewcommand\thefigure{\thesection.\arabic{figure}}
\setcounter{figure}{0}
\renewcommand{\thetable}{A\arabic{table}}
\setcounter{table}{0}

\begin{table}[h]
    \centering
    \caption{BM4 EOS parameters $\mathrm{E_0}$, $\mathrm{V_0}$, $\mathrm{B_0}$, and $\mathrm{B'_0}$ for all stable $\text{Fe}_3\text{Pt}$ configurations. $\mathrm{\Delta E}$ refers to the energy difference using the ground state as the reference. The entries are arranged in order of increasing $\mathrm{\Delta E}$.}
    \label{tab:Fe3Pt_eos}
    \begin{tabular}{c c c c c c}
        \toprule
        \multicolumn{6}{c}{\textbf{\textbf{Fe$_3$Pt}}} \\ 
        \midrule
        Configuration & $E_0$ (eV/atom) & $\Delta E$ (meV/atom) & $V_0$ (\AA$^3$/atom) & $B_0$ (GPa) & $B'_0$ \\
        \midrule
        FM  & -7.79405  & 0.00000  & 13.03605  & 175.51908  & 3.65991  \\
        SF28 & -7.78408  & 0.00998  & 12.80335  & 162.78998  & 4.18806  \\
        SF22 & -7.78014  & 0.01391  & 12.81425  & 161.86989  & 4.26958  \\
        SF4  & -7.77497  & 0.01908  & 12.91693  & 168.60494  & 2.94519  \\
        SF18 & -7.77497  & 0.01908  & 12.83218  & 162.71928  & 3.77088  \\
        SF26 & -7.76865  & 0.02541  & 12.79016  & 162.47507  & 4.24804  \\
        SF5  & -7.76683  & 0.02722  & 12.85034  & 162.60934  & 3.37632  \\
        SF25 & -7.75957  & 0.03448  & 12.77792  & 162.72771  & 3.96058  \\
        SF8  & -7.75490  & 0.03916  & 12.77759  & 167.46769  & 3.80617  \\
        SF6  & -7.75476  & 0.03929  & 12.77157  & 172.12375  & 4.08989  \\
        SF31 & -7.75203  & 0.04203  & 12.66636  & 160.94887  & 4.24354  \\
        SF11 & -7.75008  & 0.04397  & 12.76935  & 168.55766  & 4.42931  \\
        SF33 & -7.74841  & 0.04565  & 12.65939  & 157.45867  & 4.37207  \\
        SF20 & -7.74812  & 0.04594  & 12.70637  & 164.34746  & 4.50020  \\
        SF2  & -7.74544  & 0.04861  & 12.73808  & 167.95763  & 4.01685  \\
        SF10 & -7.74501  & 0.04905  & 12.73752  & 161.70944  & 3.43207  \\
        SF32 & -7.74276  & 0.05130  & 12.66074  & 157.96377  & 3.98863  \\
        SF30 & -7.74212  & 0.05193  & 12.62815  & 169.94643  & 4.42590  \\
        SF27 & -7.73976  & 0.05430  & 12.61880  & 171.52474  & 4.58797  \\
        SF34 & -7.73858  & 0.05547  & 12.62312  & 164.39756  & 4.33285  \\
        SF7  & -7.73812  & 0.05594  & 12.69141  & 172.10651  & 4.58617  \\
        SF13 & -7.73524  & 0.05882  & 12.71418  & 172.57465  & 4.61729  \\
        SF21 & -7.72981  & 0.06424  & 12.60278  & 174.60159  & 4.65484  \\
        SF17 & -7.71112  & 0.08293  & 12.60985  & 164.60052  & 4.46122  \\
        SF3  & -7.70914  & 0.08491  & 12.58432  & 167.84690  & 4.46411  \\
        \bottomrule
    \end{tabular}
\end{table}

\end{document}